\documentclass[preprint]{osa-article}

\journal{osajournal}

 \setcopyright{none}

\articletype{Research Article}

\usepackage{lineno}

\begin{document}

\title{Multi-mode Dynamics of Terahertz Quantum Cascade Lasers: spontaneous and actively induced generation of dense and harmonic coherent regimes}

\author{Carlo Silvestri,$\authormark{1}$ Xiaoqiong Qi,$\authormark{1}$ Thomas Taimre,$\authormark{2}$ Aleksandar D. Rakić,$\authormark{1,*}$}


\address{\authormark{1}School of Information Technology and Electrical Engineering, The University of Queensland, Brisbane, QLD 4072, Australia\\
\authormark{2}School of Mathematics and Physics, The University of Queensland, Brisbane, QLD 4072, Australia\\
}

\email{\authormark{*}a.rakic@uq.edu.au} 



\begin{abstract}
We present an extended study concerning the dynamics of dense and harmonic coherent regimes in quantum cascade lasers (QCLs) in a Fabry--Perot (FP) configuration emitting in the terahertz (THz) spectral region. Firstly, we study the device in free running operation, reproducing the main features of the self-generated of optical frequency combs (OFCs) and harmonic frequency combs (HFCs) in this spectral range, commenting on the points of difference from the mid-infrared region, and finding excellent agreement with the most recent experimental evidences. Then, we analyze the THz-QCL dynamics under radiofrequency (RF) injection, with a focus on the effect of the modulation of the current on the degree of locking of the system, and we perform a systematic investigation aimed to provide a procedure for the generation of train of pulses with short duration and high contrast. Furthermore, we extend our study to the generation of sequences of pulses with repetition frequency which is a multiple of the free-spectral range (FSR) of the laser cavity, reproducing harmonic mode-locking (HML) of the laser, a method which has been recently demontrated in experiments with THz-QCLs, and the results of which are particulary promising for a large stream of applications, ranging from optical communication to imaging.
\end{abstract}
\section{Introduction}
Since the first demonstration in 2002\cite{Kohler}, quantum cascade lasers (QCLs) emitting in the terahertz (THz) spectral region have attracted an increasing interest and growing technological development. In particular, the experimental discovery that these devices can spontaneously emit optical frequency combs (OFCs) \cite{Wienold14}, \cite{Burghoff14}, multi-mode coherent regimes defined as set of equally spaced optical lines with low phase and amplitude noise \cite{PiccardoReview}, has made them ideal for applications in the field of metrology, sensing, high-precision spectroscopy and dual comb spectroscopy, considering that many molecules and chemical compounds present absorption lines in the THz range \cite{revFaist}, \cite{Villaresdualcomb}, \cite{Consolinodualcomb}, \cite{metrology}. Furthermore, not only dense OFCs have been found experimentally, where by ''dense'' we mean regimes characterized by a fundamental spacing between the optical lines in the optical spectrum corresponding to the free-spectral range (FSR) of the laser cavity, but also self-starting harmonic frequency combs (HFCs), i.e. analogous coherent regimes with spacing a multiple of the FSR, have been experimentally found. In particular, these two types of coherent regime emitted from the same device alternate for different values of the DC bias current, as it has been observed in both mid-infrared (mid-IR) \cite{Kazakov2017}, \cite{PiccardoReview} and THz \cite{HFCTHZ}, \cite{ForrerHFCTHZ}, and in general, coherent regimes are alternated with irregular unlocked states by sweeping the bias current \cite{Li15}.\\
If we look at the OFCs in the time domain, THz-QCL OFCs present differences with respect to the mid-IR region. First of all, while in the mid-IR a regular and identical repetition of power spikes separated by a quasi-constant background has been reported in the reconstructed power versus time traces obtained by adopting the SWIFTS technique \cite{OpticaFaist}, in the THz region recent measurements based on a combination of dual-comb multi-heterodyne detection and Fourier-transform analysis have shown a power trace characterized by a coherent emission of multi-peaked structures, without proper single spikes \cite{Cappelli}. Secondly, these OFC regimes are characterized by a frequency modulation (FM) behaviour, but although a clear linear chirp for the instantaneous frequency has been observed in the mid-IR \cite{OpticaFaist}, for THz QCL experimental measurements have not shown this trend as clearly \cite{Cappelli}.\\
Furthermore, we mention that recent real-time measurements based on intracavity self detection in long-cavity ($15 \mathrm{mm}$) THz-QCL have also reported the occurrence of coherent population pulsations with fundamental and harmonic repetition time, associated to OFC emission \cite{Li22}.\\
Another way to achieve coherent emission from QCLs is the performance of active mode-locking by modulating the bias current in a portion of the laser cavity at the beatnote (BN) frequency through the injection of a RF signal, firstly demonstrated for mid-IR QCLs \cite{Paiella} and later for THz-QCLs \cite{Gellie1}, \cite{Barbieri_2011}. The active-mode locking induces the generation of a train of pulses, with a repetition time corresponding to the roundtrip time (RT) of the laser cavity. More recently, the first demonstration of harmonic mode-locking (HML) has been presented, with the generation of multiple pulses per roundtrip, by modulating the QCL at the multiple of the BN frequency \cite{Dhillon1}.\\
From the theoretical point of view several works have been done in order to describe the spontaneous and the active mode-locked generation of coherent regimes in QCLs, and more recently also in presence of optical injection \cite{Pratinano}, \cite{Unifying}. The initial theoretical approaches were based on a classical set of Maxwell--Bloch equations \cite{Wang15}, \cite{Villarestheory}, \cite{Boiko}, \cite{Jirauschek}, not including the mode-amplitude coupling for the electric field provided by the linewidth enhancement factor (LEF, also called $\alpha$ factor). Therefore, for the study of self-starting comb regimes, the close to threshold multi-mode instability was triggered only from spatial hole burning (SHB). The following introduction of the LEF in full \cite{Columbo18}, \cite{Silvestri20} and reduced \cite{PRLOpacak} models, has brought to succesfully reproduce the main features of self-generated OFC, i.e. the  coexistence of FM and AM behaviour of OFC in QCL, and provide a reproduction of the linear chirp behaviour observed in Fabry--Perot (FP) mid-IR QCLs. Furtherly, the development of a mean field theory approach has introduced an analytical description of the linear chirp characterizing spontaneous OFCs \cite{Burghoff20}, \cite{Humbard22}.\\
In this work we present an extensive study of dense and harmonic coherent regimes in THz-QCLs both, spontaneously generated and induced with active mode-locking, by adopting the effective semiconductor Maxwell--Bloch equations (ESMBEs) for the FP configuration \cite{Silvestri20}.\\
In Section \ref{model} we introduce the model, which has the peculiar feature to provide an inclusion of the SHB and encompasses self-consistently the main properties of semiconductor materials, such as the asymmetric gain/refractive index spectra (leading to a non null $\alpha$ factor), the dependence of the optical response of the active medium on the density of carriers, and the inclusion of nonlinearities at each order (four-wave mixing, Kerr effect, and so on), resulting therefore a suitable platform for the investigation of the multi-mode operation of these devices.\\
Section \ref{Sec1} is dedicated to the numerical study of the laser in free running operation with a focus on the coherent multi-mode dynamics of the device. We are capable to successfully reproduce the experimentally demonstrated features of self-starting OFC regimes in the THz region, such as: the regular repetition of multi-peaked structures in the power trace and the FM behaviour which is not a clear linear chirp of the instantaneous frequency, where both these characteristics allow a distinction between THz and mid-IR OFCs; alternating locked and unlocked states; the spontaneous generation of HFC regimes; the pulsations of the density of carriers occurring with repetition rate the roundtrip time of the QCL cavity (RT) and twice the roundtrip time (2RT). In particular, to the best of our knowledge, we show for the first time the reproduction of self-generated HFC regimes emitted by a THz QCL in FP configuration by exploiting a model based on a single electronic transition, in contrast with a previous theoretical study which stated that this phenomenon is strongly favoured by considering two asymmetric transitions \cite{ForrerHFCTHZ}. \\
In Section \ref{RFinj} we study the emission of coherent regimes under RF injection, by introducing the modulation of the current in our model. Firstly, we show the numerical results related to a modulation performed to the entire cavity. It is well known that a train of single peaked pulses can not be achieved in this configuration, and our investigation is focused on how the RF injection modifies the degree of coherence of the OFC and HFC states obtained in free running operation. We find out that three different regimes occur: for low values of the modulation depth, the coherence of the system is damaged, resulting in slow modulations of the emitted power; for higher values of the modulation depth, the systems re-enters a regime of locking; by further increasing the modulation depth, we report a transition of the laser dynamics into an unlocked region, characterized by fast power modulations, and the threshold value of the modulation depth where this transition occurs depends on the central value of the bias current. Furthermore, we observed that HFC regimes falls into dense states for a consistent value of the amplitude after application of RF injection at the same harmonic of the repetition frequency of the considered HFC. Secondly, we study the dynamics of the laser when a section of the cavity is modulated, and a coherent regular train of pulses is obtained. In particular, we examine the impact of each RF parameter (length of the modulated section, DC value of the current in the unmodulated section, modulation depth, modulation frequency) on the contrast, peak power and pulse duration, in order to have a prescription for the achievement of the shortest pulses with higher emitted peak power and maximum contrast. Thirdly, we present a numerical study concerning the harmonic mode-locking of a THz-QCL, showing the generation of train of multiple short pulses each RT time, and pushing our analysis to the fifth harmonics, where a pulse duration on the ps scale with a repetition time of 10ps are achieved.\\
Finally, Section \ref{conclusion} draws the conclusions of the work.

\section{Model: Effective Semiconductor Maxwell--Bloch Equations for a Quantum Cascade Laser in Fabry-Perot configuration}\label{model}
For the study of the multi--mode dynamics of a QCL in FP configuration emitting in the THz region, we adopt in this work the ESMBEs, originally introduced in \cite{Columbo18} for the study of a QCL in the ring configuration, and developed for the FP resonator in \cite{Silvestri20}, for the analysis of a mid-IR QCL. This model encompasses the optical susceptibility of a semiconductor material, which properly describes the medium of a QCL, and accounts for SHB, including the carriers grating which characterizes a FP resonator.\\
We consider a FP-QCL with cavity length of a few millimeters, and we assume that the electric field is expressed as a superposition of two counterpropagating fields inside the resonator:
\begin{equation}
E(z,t)=\frac{1}{2}[E(z,t)^+\exp(-ik_0z+i\omega_0 t)+E(z,t)^-\exp(+ik_0z+i\omega_0 t)+c.c.] \label{el1}\\
\end{equation}
where $E^+(z,t)$, $E^-(z,t)$ are respectively the slowly varying envelopes for the forward and backward fields inside the laser cavity, $\omega_0$ (which is used as the reference frequency) is the mode of the cold cavity closest to the peak of the gain, and $k_0$ is the corresponding wavenumber. The polarization is given by:
\begin{equation}
P(z,t)=\frac{1}{2}[P_0(z,t)\exp(+i\omega_0 t)+c.c.], \label{pol1}
\end{equation}
where $P_0(z,t)$ is the envelope of the polarization. By using a multiple scale approach it is possible to express the polarization envelope $P_0(z,t)$ and the carrier density $N(z,t)$ by exploiting truncated Fourier series expansions of the spatial variation at the wavelength \cite{Silvestri20}, \cite{Bardella17}, \cite{lugiato_prati_brambilla_2015}:
\begin{eqnarray}
P_0(z,t)&=&P_0^+(z,t)\exp(-ik_0z)+P_0^-(z,t)\exp(+ik_0z),\label{Pfour}\\
N(z,t)&=&N_0(z,t)+N_1^+(z,t)\exp\left(-2ik_0z\right)+N_1^-(z,t)\exp\left(+2ik_0z\right),\label{nfour}
\end{eqnarray}
where $P_0^+$, $P_0^-$ are respectively the forward and backward terms of the polarization envelope, $N_0$ is the zero-order density of carriers and $N_1^+$, $N_1^-$ are the first order terms of the carrier density and represent the carrier grating due to SHB occurring in the FP resonator. Since $N$ is a real-valued quantity, $N_1^+=$ $N_1^{-*}$.\\
\begin{figure}[t]
\centering
\includegraphics[width=0.60\textwidth]{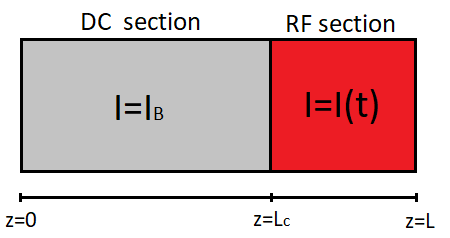}
\caption{Scheme of the cavity under RF injection. $L_\mathrm{C}$ is the length of the portion of the cavity with constant value $I_\mathrm{B}$ of the bias current. Between z=$L_\mathrm{C}$ and z=$L$ the current is a function of the time $I(t)$.}
\label{schememod}
\end{figure}
By exploiting these hypotheses it is possible to write (for the details about the derivation of the equations, see \cite{Silvestri20}) the ESMBEs for a QCL in FP configuration:
\begin{eqnarray}
\frac{\partial E^+}{\partial z}+ \frac{1}{v}\frac{\partial E^+}{\partial t} &=& -\frac{\alpha_\mathrm{L}}{2}E^++g P_0^+ ,\label{el+}\\
-\frac{\partial E^-}{\partial z}+ \frac{1}{v}\frac{\partial E^-}{\partial t} &=& -\frac{\alpha_\mathrm{L}}{2}E^-+g P_0^- ,\label{el-}\\
\frac{\partial P_0^+}{\partial t}&=&\frac{(1+i\alpha)}{\tau_\mathrm{d}}\left[-P_0^++if_0\varepsilon_0\varepsilon_\mathrm{b}\left(1+i\alpha\right)\left(N_0E^{+}+N_1^+E^-\right) \right], \label{P+}\\
\frac{\partial P_0^-}{\partial t}&=&\frac{(1+i\alpha)}{\tau_\mathrm{d}}\left[-P_0^-+if_0\varepsilon_0\varepsilon_\mathrm{b}\left(1+i\alpha\right)\left(N_0E^{-}+N_1^-E^+\right) \right], \label{P-}\\
\frac{\partial N_0}{\partial t}&=&\frac{I}{eV}-\frac{N_0}{\tau_\mathrm{e}}+\frac{i}{4\hbar}\left[E^{+*}P_0^++E^{-*}P_0^--E^{+}P_0^{+*}-E^{-}P_0^{-*}\right], \label{N0}\\
\frac{\partial N_1^+}{\partial t}&=&-\frac{N_1^+}{\tau_\mathrm{e}}+\frac{i}{4\hbar}\left[E^{-*}P_0^+-E^{+}P_0^{-*}\right]. \label{N1}
\end{eqnarray}
where $v$ is the group velocity, $\alpha_\mathrm{L}$ is the loss term, $\alpha$ is the linewidth enhancement factor (LEF), $\tau_\mathrm{d}$ is the polarization dephasing time, $f_0$ is the differential gain, $\varepsilon_0$ is the vacuum dielectric constant, $\varepsilon_\mathrm{b}$ is the relative dielectric constant of the active medium, $I$ is the pump current, $V$ is the volume of the active medium, $\tau_\mathrm{e}$ is the carrier lifetime, and the coefficient $g$ is given by:
\begin{equation}
g=\frac{-i\omega_0N_\mathrm{p}\Gamma_\mathrm{c}}{2\varepsilon_0nc} \, , \label{g}
\end{equation}
where $N_p$ is the number of stages of cascading structure,  $\Gamma_c$ is the confinement factor of the radiation, $c$ is the speed of light in a vacuum, and $n$ is the effective background refractive index of the medium. The proper boundary conditions characterizing the FP resonator are taken into account:
\begin{eqnarray}
E^-(L,t)&=&\sqrt{R}E^+(L,t),\label{bc1}\\
E^+(0,t)&=&\sqrt{R}E^-(0,t),\label{bc2}
\end{eqnarray}
where $R$ is the reflectivity of each mirror and $L$ is the length of the cavity. For the further analysis of the multimode dynamics of the laser, we introduce the quantity $\delta_{\mathrm{hom}}=\frac{1}{\pi \tau_\mathrm{d}}$, which represents the full width at half-maximum (FWHM) of the gain spectrum in the limit of two level system (i.e. where we consider $\alpha \ll 1$).\\
In this work we are interested in considering not only the laser in free running operation, but also a time dependent bias current $I$, in order to describe the multi-mode dynamics of the laser under RF injection. Furthermore, we also want to investigate cases where a portion of the cavity experiences a modulation of the current, while the remaining part is set in DC operation, as shown in the scheme presented in Fig.~\ref{schememod}. To model this configuration, we introduce a general function which describes the behaviour of $I$ at each point of the laser cavity:
\begin{eqnarray}
I(z,t)=
\left\{
	\begin{array}{ll}
		I_\mathrm{B}  & \mbox{if } 0 \leq z \leq L_\mathrm{C} \\
		I_0+I_\mathrm{A} \cos(2\pi\Omega_\mathrm{M}t) & \mbox{if } L_\mathrm{C} < z \leq L
	\end{array}
\right.\label{RFformula}
\end{eqnarray}
where $I_\mathrm{B}$ is the DC value of the bias current in the unmodulated part, $I_0$ and $I_\mathrm{A}$ are respectively the central value and modulation amplitude of the bias current in the modulated portion and $\Omega_\mathrm{M}$ is the modulation frequency. For further convenience, we introduce the parameter
\begin{equation}
p=1-\frac{L_\mathrm{C}}{L} \label{pparam}
\end{equation}
which represents the fraction of the cavity which is subjected to RF injection.
\section{Free running Operation: numerical results}\label{Sec1}
In this section we study the multi-mode dynamics of the QCL in free running operation, which corresponds to the case $p=0$, with the goal to investigate the self-generation of coherent regimes in the THz region, and reproduce the most recent and significant experimental outcomes with our numerical results \cite{Cappelli}, \cite{Li15}, \cite{Li22}. For this purpose, we solve the ESMBEs (\ref{el+})--(\ref{N1}) numerically with the boundary conditions (\ref{bc1}) and (\ref{bc2}) by choosing reasonable values of the parameters of the model for the description of a THz-QCL, as shown in Table \ref{Table1}. The code used for the integration of the ESMBEs exploits a travelling wave time domain (TWTD) algorithm, based on a finite difference scheme with discretization both in time and space \cite{Silvestri20}, \cite{Bardella17}. The same code will also be used to obtain the results presented in the other sections of this work.

\begin{table}
\centering
\begin{tabular}{|c | c | c | c | c | c | c | c | c | c |} 
 \hline
 $L(\mathrm{mm})$ & $n$ & $R$ & $\tau_\mathrm{e} (\mathrm{ps})$ & $\Gamma_c$ & $f_0 (\mu\mathrm{m}^3)$ & $V (\mu\mathrm{m}^3)$ & $N_p$ & $\alpha$ & $\nu_0 (\mathrm{THz})$ \\ [0.5ex] 
\hline
 $2$ & $3.6$ & $0.3$ & $5$ & $0.13$ & $7\cdot10^{-5}$ & $3.6\cdot10^{6}$ & $90$ & $-0.1$ & $3$\\ 
\hline
\end{tabular}
\caption{\label{Table1}Typical parameters of THz-QCL.}
\end{table}
\begin{figure}[h]
\centering
\includegraphics[width=1\textwidth]{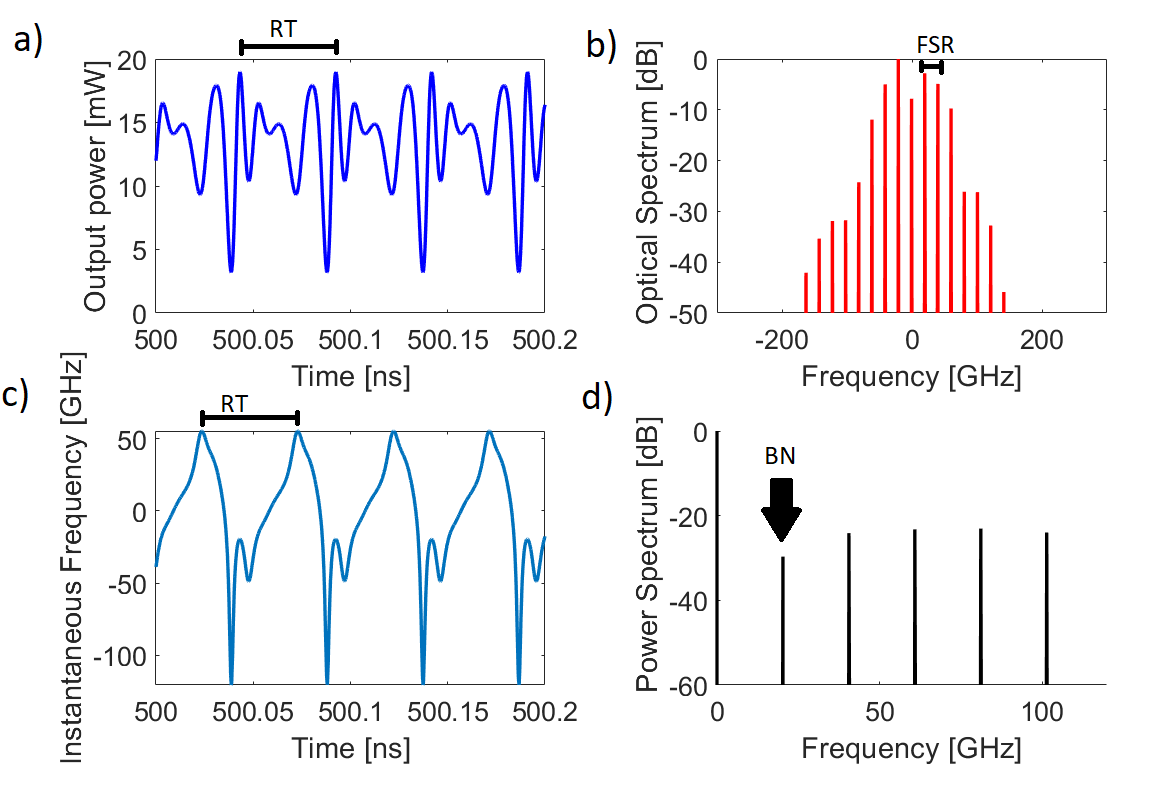}
\caption{Dense spontaneously generated OFC for $I/I_\mathrm{thr}=1.54$, $\delta_\mathrm{hom}=0.32$THz and the parameters as in Table \ref{Table1}. a) Output power as a function of the time. The temporal distance between two consecutive maxima corresponds to the roundtrip time (RT) of the resonator. b) Optical spectrum of the radiation. The separation between two adjacent peaks is the free spectral range (FSR) of the laser cavity. c) Temporal evolution of the instantaneous frequency. d) Power spectrum, exhibiting peaks at the fundamental beatnote (BN) frequency and at its multiples.}
\label{OFCdense}
\end{figure}
The aim of this study is the characterization of the multi-mode regimes obtained by simulating the laser with the ESMBEs, with a focus on the identification of optical frequency comb (OFC) regimes. In order to classify a simulated multi-mode regime, we calculate the comb indicators $M_{\sigma_P}$ and $M_{\Delta\Phi}$, previously introduced in \cite{Silvestri20}, which are respectively an estimation of the amplitude and phase noise of the considered dynamical regime. These indicators are obtained by performing a post-processing analysis of the numerically calculated optical field, which is easily accessible by the TDTW model; first, each line $q$ of the output optical field in the $-10\mathrm{dB}$ spectral bandwidth is filtered, and then the temporal dynamics of the power of each line ($P_q(t), q=1,...,N_{10}$) is retrieved, and the temporal evolution of the phase difference two adjacent modes ($\Delta\Phi_q(t), q=1,...,N_{10}$), where $N_{10}$ is the number of modes included in the $-$10dB bandwidth from the maximum in the optical spectrum. The definition of these indicators is given by:
\begin{equation}
M_{\sigma_{P}}=\frac{1}{N_{10}}\sum_{q=1}^{N_{10}}{\sigma_{P_q}}\quad , \quad M_{\Delta\Phi}=\frac{1}{N_{10}}\sum_{q=1}^{N_{10}}{\sigma_{\Delta\Phi_q}},\label{MsigmaMPhi}
\end{equation}
where:
\begin{equation}
\mu_{P_q}=\left\langle P_q(t)\right\rangle \quad , \quad
\mu_{\Delta\Phi_q}=\left\langle\Delta\Phi_q(t)\right\rangle,
\end{equation}
\begin{equation}
\sigma_{P_q}=\sqrt{\left\langle\left(P_q(t)-\mu_{P_q}\right)^2\right\rangle} \quad , \quad
\sigma_{\Delta\Phi_q}=\sqrt{\left\langle\left(\Delta\Phi_q(t)-\mu_{\Delta\Phi_q}\right)^2\right\rangle}.
\end{equation}
The symbol $\left\langle\right\rangle$ indicates the average taken over the time. OFC regimes are identified by values of these indicators which are ideally zero. Since in our simulations there are residual fluctuations  of modal power and intermodal phase difference, we define a numerically simulated dynamical regime as an OFC if $M_{\sigma_{P}}<2\cdot10^{-2} \mathrm{mW}$ and $M_{\Delta \Phi}<2\cdot10^{-2} \mathrm{rad}$.\\
At this point, we can summarize the procedure that we follow in our numerical study for the free running laser with the following steps:
\begin{itemize}
  \item  we fix the value of $\delta_\mathrm{hom}$, which corresponds to fix the dephasing time of the polarization, and the gain bandwidth of our device;
  \item we consider a set of values of the bias current $I$ between the threshold current $I_\mathrm{thr}$ and 3$I_\mathrm{thr}$  with fixed  current step $\Delta I=$0.08$I_\mathrm{thr}$;
  \item we perform a dynamical simulation by integrating the ESMBEs at each considered value of the bias current;
  \item we classify each simulated dynamical regime by using the indicators $M_{\sigma_P}$ and $M_{\Delta\Phi}$ defined by Eq. (\ref{MsigmaMPhi}), identifying locked and unlocked regimes.
\end{itemize}
\begin{figure}[t]
\centering
\includegraphics[width=1\textwidth]{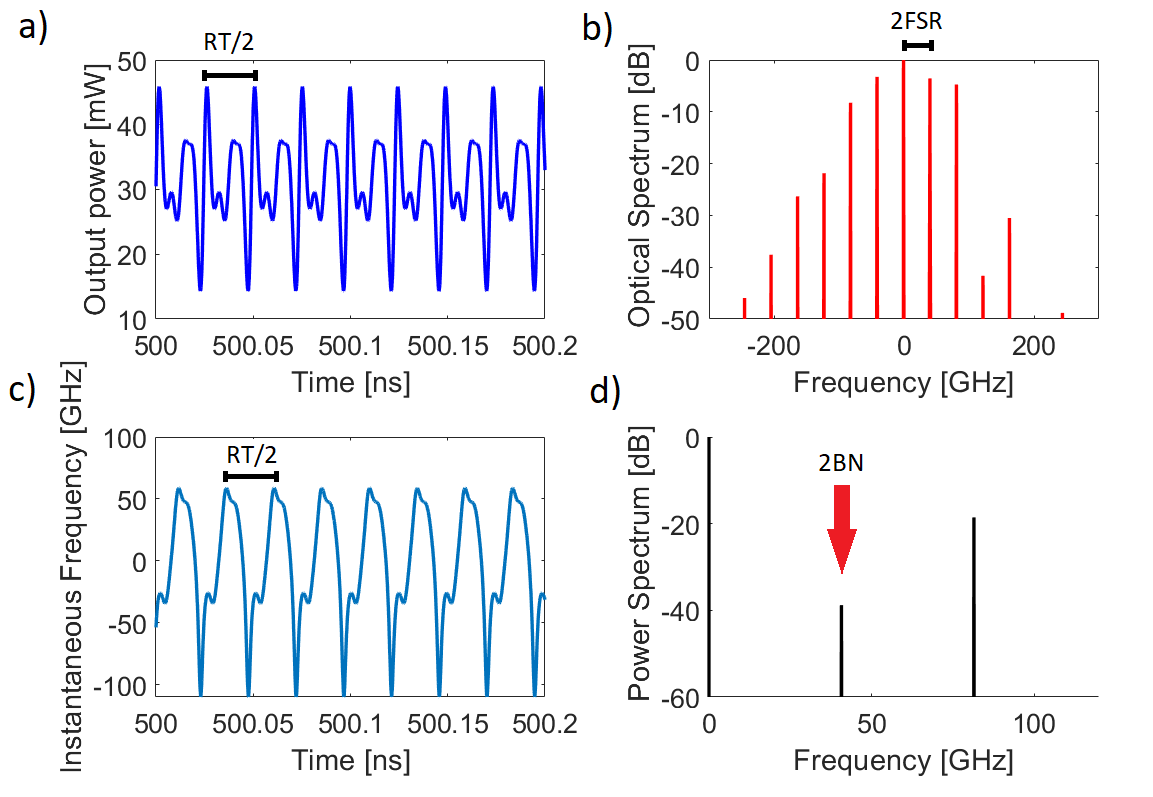}
\caption{Self-generated harmonic frequency comb for $I/I_\mathrm{thr}=2.15$, $\delta_\mathrm{hom}=0.32$THz and the parameters as in Table \ref{Table1}. a) Output power as a function of the time. The temporal distance between two consecutive maxima corresponds to half the roundtrip time (RT/2) of the resonator. b) Optical spectrum of the radiation. The separation between two adjacent peaks is twice the free spectral range (2FSR) of the laser cavity. c) Temporal evolution of the instantaneous frequency. d) Power spectrum, exhibiting peaks at twice the beatnote frequency (2BN) and at its multiples.}
\label{HFC}
\end{figure}
We first present the results for $\delta_\mathrm{hom}$=0.32THz, with $I_\mathrm{thr}$=650mA and the other parameters as in Table 1.\\
An example of dense OFC regime, found at $I$=1.54$I_\mathrm{thr}$, is shown in Fig.~\ref{OFCdense}. The temporal evolution of the emitted power (Fig.~\ref{OFCdense}a) is characterized by a regular repetition of a multi-peaked structure each roundtrip time, while the instantaneous frequency (Fig.~\ref{OFCdense}c) presents a quasi-linear chirp trend, but not as clear as it has been observed in the mid-IR region \cite{OpticaFaist}, showing a similarity with recent experimental observations of OFC regimes in THz-QCLs \cite{Cappelli}. The optical spectrum (Fig.~\ref{OFCdense}b) is characterized by 6 modes in the first decade, and the power spectrum presents lines at the fundamental beatnote and at its multiples.\\
Another remarkable case of locked regime has been obtained for $I$=2.15$I_\mathrm{thr}$, and it is presented in Fig.~\ref{HFC}. 
The power trace, shown in Fig.~\ref{HFC}a, is characterized by an identical repetition of a structure composed by a main peak and a secondary bump and it differs from the OFC presented in Fig.~\ref{OFCdense} because of its repetition time, which is here half the round-trip time (RT/2). Coherently, the instantaneous frequency (Fig.~\ref{HFC}c) regularly repeats itself on the same timescale, and the optical spectrum (Fig.~\ref{HFC}b) presents a spacing between two consecutive optical lines corresponding to twice the free spectral range (2FSR). Finally, in the power spectrum (Fig.~\ref{HFC}d) we do not observe any peak at the fundamental beatnote (BN), but at twice the beatnote (2BN) and its multiples. We can, therefore, conclude that the considered coherent regime is a self-starting harmonic frequency comb (HFC), a type of dynamics which has been experimentally measured in THZ-QCLs \cite{Forrer2020}, \cite{ForrerHFCTHZ}.\\
\begin{figure}[t]
\centering
\includegraphics[width=0.8\textwidth]{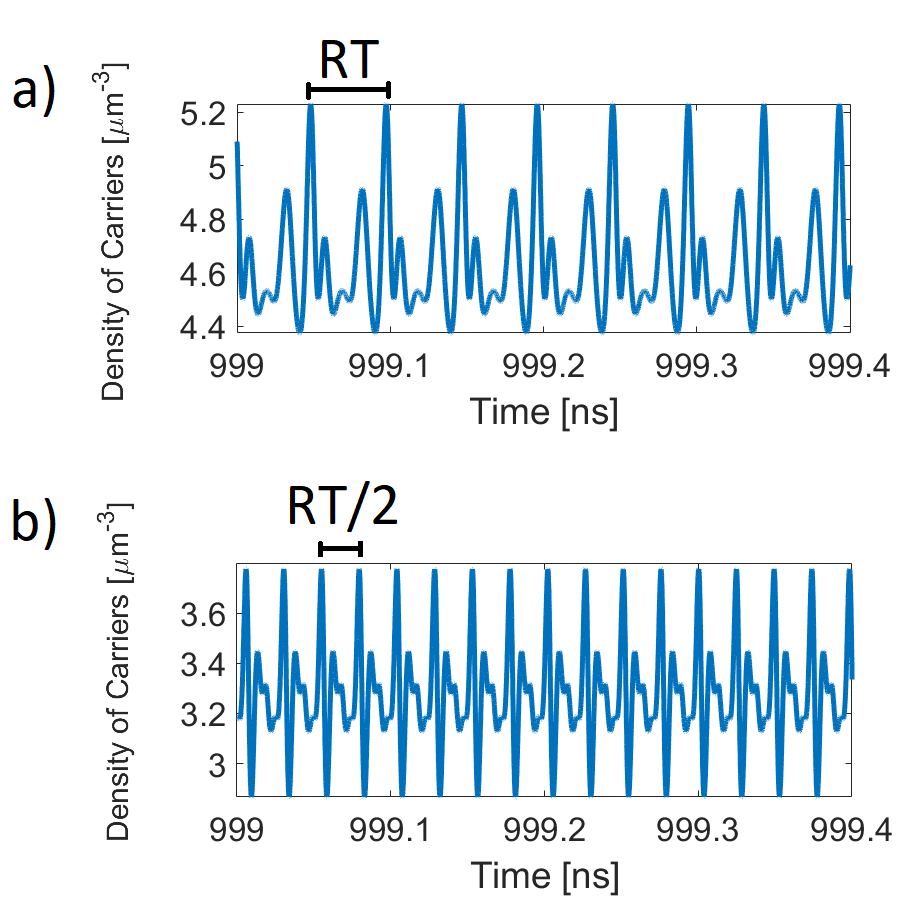}
\caption{Temporal evolution of the density of carriers at the right facet of the QCL $z=L$, for the OFC (a) and HFC (b) regimes shown respectively in Fig.~\ref{OFCdense} and Fig.~\ref{HFC}. We observe sequences of spikes with repetition time corresponding to the roundtrip time, and twice the roundtrip time, respectively.}
\label{carrierpulses}
\end{figure} 
Furthermore,  for the OFC and HFC regimes just presented respectively in Fig.~\ref{OFCdense} and \ref{HFC}, we can look at the dynamics of the density carriers, and we find a regular repetition of carrier spikes, sorrounded by low secondary peaks, with repetition time RT and 2RT respectively. These results are in agreement with the experimental findings presented in \cite{Li22}, where exactly carrier pulsations with repetition time RT and 2RT have been demonstrated with a real-time measurement technique performed on a THz-QCL.\\
\begin{figure}[!h]
\centering
\includegraphics[width=1\textwidth]{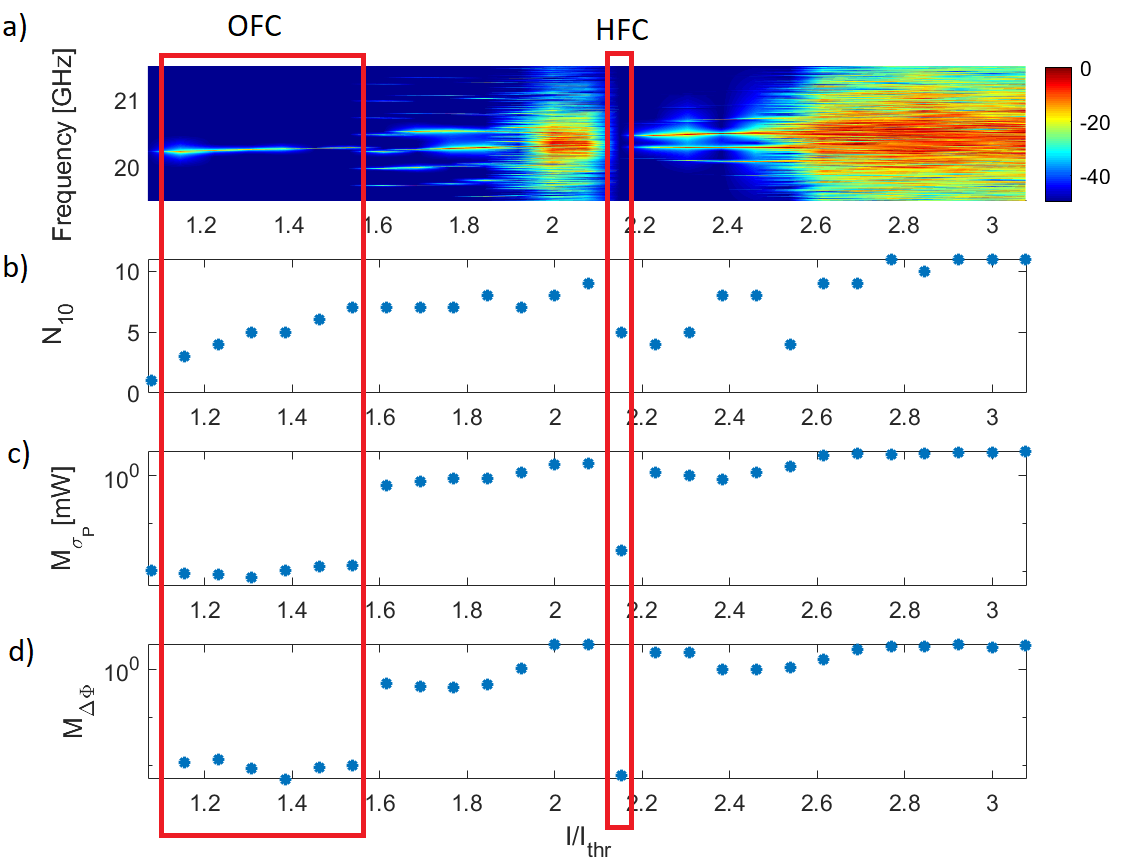}
\caption{Results obtained by performing a bias current scan between $I_\mathrm{thr}$ to $3.1I_\mathrm{thr}$ with current step $0.07I_\mathrm{thr}$  for $\delta_\mathrm{hom}$=0.32THz and parameters as in Table \ref{Table1}. a) Map showing the first BN in the power spectrum (color scale normalized to the maximum for each current value; log scale); b) number of modes in -10dB bandwidth of the optical spectrum; comb indicators for amplitude (c) and  phase (d) noise, defined in Eq. (\ref{MsigmaMPhi}). We indicate with OFC the region where only dense comb regimes have been found, and with HFC the region with harmonic combs.}
\label{currentscan}
\end{figure} 
If we consider the dynamics of the simulated QCL along all the considered current interval, which is presented in Fig.~\ref{currentscan}, we can notice that for the lowest value of $I$ the laser presents a single-mode emission, and then from $I$=1.15$I_\mathrm{thr}$ the multi-mode operation occurs. Between $I$=1.15$I_\mathrm{thr}$ and $I$=1.54$I_\mathrm{thr}$ a narrow beatnote line is observable in the RF spectrum (see Fig.~\ref{currentscan}a), which corresponds to low values of $M_{\sigma_P}$ and $M_{\Delta\Phi}$, as noticeable in Fig.~4a)--b): we have therefore a first window characterized by locked regimes, which is higlighted in red color in Fig.~\ref{currentscan}. An unlocked dynamics rises at $I$=1.6$I_\mathrm{thr}$, denoted by the presence of several distinguishable peaks in the BN map up to $I$=1.9$I_\mathrm{thr}$, which corresponds to the occurrence of amplitude modulations in the power temporal evolution, and then by a large broadened BN between $I$=1.9$I_\mathrm{thr}$ and $I$=2.1$I_\mathrm{thr}$, which is associated with chaotic regimes. At $I$=2.15$I_\mathrm{thr}$ the first BN disappears and we found a coherent state which is the HFC shown in Fig.~\ref{HFC}. We performed a dense bias current scan around this point, and we identified a narrow HFC window 40mA large. Finally, for $I>$2.15$I_\mathrm{thr}$ only unlocked regimes are found. These results reproduce the alternance between locked and unlocked regimes reported in the experimental literature for THz QCLs \cite{Li15}, with the occurrence of self-starting HFCs. Therefore, we can observe that the ESMBEs for FP-QCLs are able to reproduce the spontaneous generation not only of dense OFC states, but also self-starting coherent harmonic regimes, as reported in the experiments \cite{Kazakov2017}, \cite{ForrerHFCTHZ}.\\
The same analysis presented for the case $\delta_{\mathrm{hom}}=$0.32THz has been performed for $\delta_{\mathrm{hom}}=$0.16THz and $\delta_{\mathrm{hom}}=$0.48THz, and all the parameters as in Table \ref{Table1}.
\begin{figure}[t]
\centering
\includegraphics[width=1\textwidth]{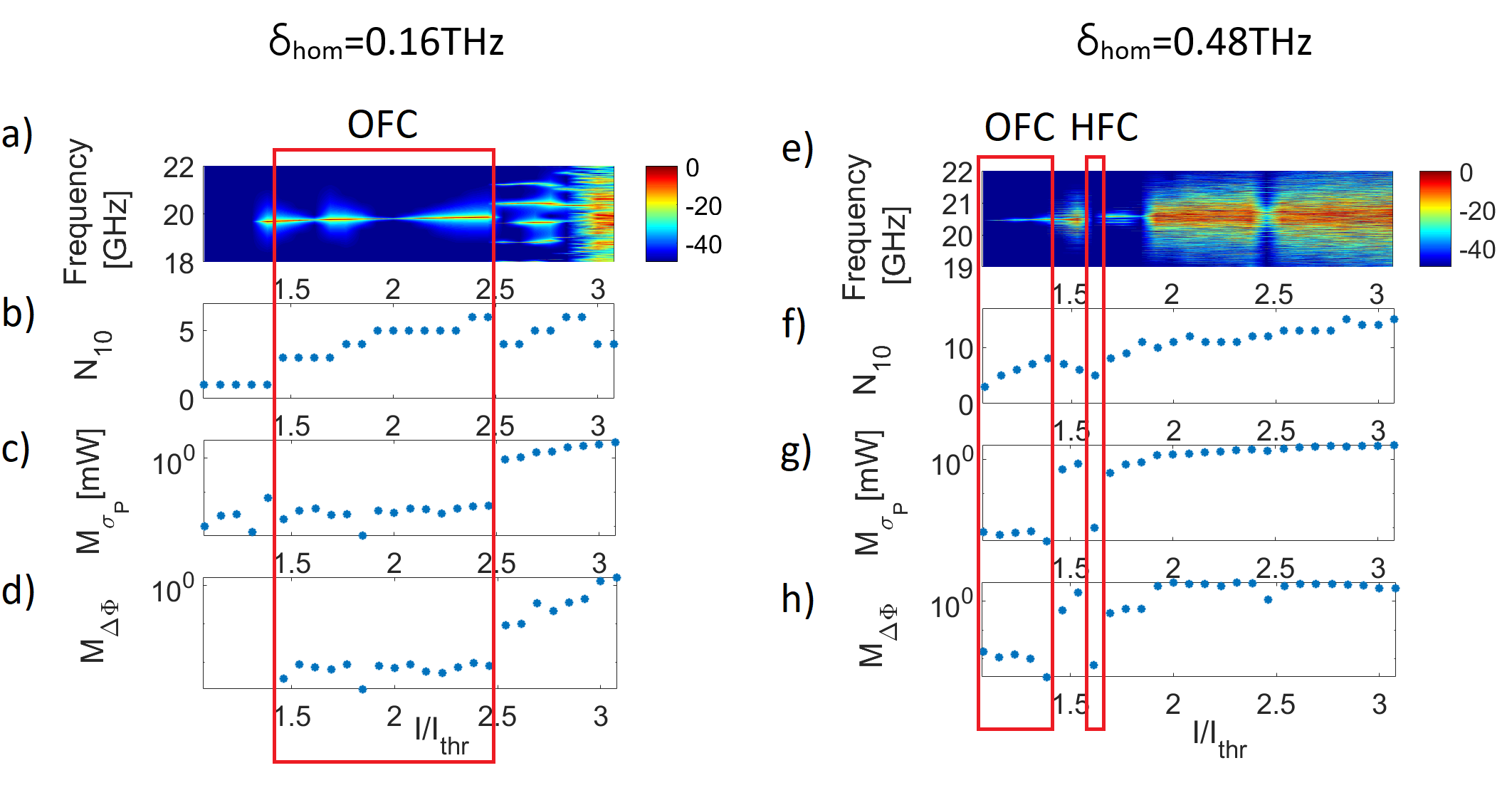}
\caption{Results obtained by performing a bias current scan between $I_\mathrm{thr}$ to $3.1I_\mathrm{thr}$ for $\delta_{\mathrm{hom}}=$0.16THz (left) and $\delta_{\mathrm{hom}}=$0.48THz (right). Other parameters as in Table \ref{Table1}. a)--e) Map showing the first BN in the power spectrum (color scale normalized to the maximum for each current value; log scale); b)--f) number of modes in -10dB bandwidth of the optical spectrum; comb indicators for amplitude (c)--(g) and (d)--(h) phase noise, defined in Eq. (\ref{MsigmaMPhi}). We indicate with OFC and highlight in red the region where only dense comb regimes have been found.}
\label{currentscan_diff}
\end{figure} 
As can be noticed in Fig.~\ref{currentscan_diff}, for the lower value of gain bandwidth we report a large comb region, and the maximum number of locked modes in the first decade of the optical spectrum is 6. The OFC region is followed by an unlocked window. Conversely, for $\delta_\mathrm{hom}=$0.48THz we observe the onset of the multi-mode dynamics very close to threshold, with a narrower comb region respect to the other studied cases, followed by an unlocked window and a narrow HFC region, analogously with the case $\delta_\mathrm{hom}=$0.48THz previosuly presented in Fig. \ref{currentscan}. The evolution of the dynamical scenario with the gain bandwidth here reported for THz QCL presents close similarity with the evidences from the previously obtained numerical results concerning the mid-IR range \cite{Silvestri20}.\\


\section{Effect of RF modulation: numerical results}\label{RFinj}
\subsection{RF injection in the entire cavity}
We want to study how the RF injection affects the OFC and HFC regimes obtained in the free running operation, which have been presented in the Sec. \ref{Sec1}. In particular, we want to answer the following questions:
\begin{itemize}
    \item How does the modulation of the current modify the temporal evolution of the power of an OFC regime obtained in free running operation?
    \item Does the modulation of the current keep the locking degree of these OFCs?
    \item How does the modulation depth $I_A$ influence the degree of locking?
    \end{itemize}
In order to answer these questions we first consider the case $p=1$, which corresponds to the modulation of the current implemented in the entire cavity of the QCL. Therefore, for every point in the laser cavity the bias current is expressed as:
\begin{eqnarray}
I(t)=
		I_0+I_\mathrm{A}\cos(2\pi\Omega_\mathrm{M}t)\label{Ient}.
\end{eqnarray}
\begin{figure}[t]
\centering
\includegraphics[width=1\textwidth]{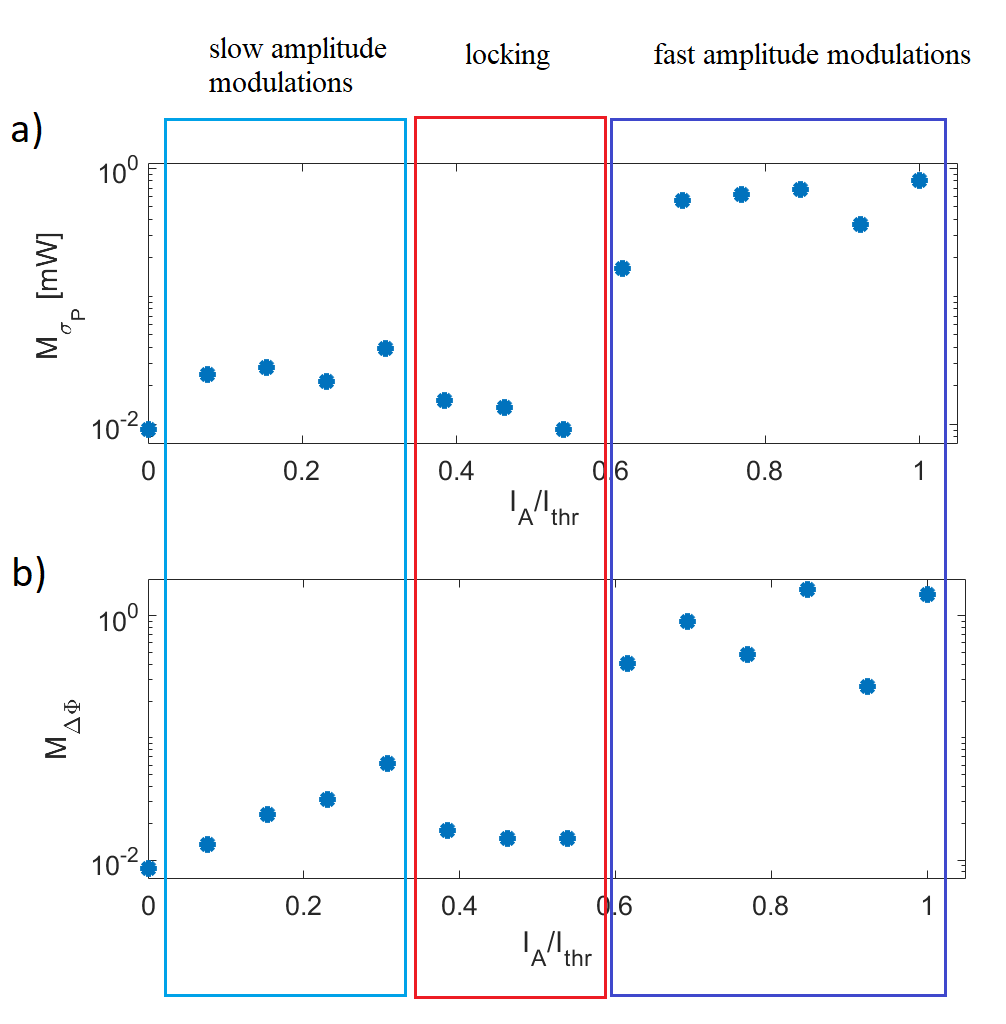}
\caption{RF injection in the entire cavity: indicators for amplitude (a) and phase (b) noise for different simulated regimes obtained by integrating the ESMBEs with a modulate bias current expressed by Eq. \ref{RF_entire}, for $I_0=$1.54$I_\mathrm{thr}$, $\Omega_\mathrm{M}$ equals to the BN frequency extracted from the free running simulation at $I$=$I_0$, $\delta_\mathrm{hom}=$0.32THz, other parameters as in Table \ref{Table1}, and different values of $I_A$ between 0.08$I_\mathrm{thr}$ and $I_\mathrm{thr}$ with step $\Delta I_\mathrm{A}=$0.08$I_\mathrm{thr}$. The points for $I_\mathrm{A}=$0 corresponds to the free running case shown in Fig.~\ref{OFCdense}.}
\label{RF_entire}
\end{figure}
In our numerical investigation we choose the parameters as in Table \ref{Table1} and we fix $\delta_\mathrm{hom}=$0.32THz. Then we determine the BN frequency $f_\mathrm{BN}$ from a simulation of total duration 1$\mu\mathrm{s}$ performed at a fixed value of the bias current $I_0$ in free running operation, and we implement the extracted value of the BN frequency as modulation frequency $\Omega_\mathrm{M}$ in Eq. (\ref{Ient}). We remark here that $f_{BN}$ is different at each value of the bias current, since in our model the optical susceptivity, and therefore the refractive index, is a quantity dependent on the density of carriers $N$, which in turns depends on the bias current through Eq. (\ref{N0}); therefore the FSR depends on $I$ as well.
\begin{figure}[t]
\centering
\includegraphics[width=1\textwidth]{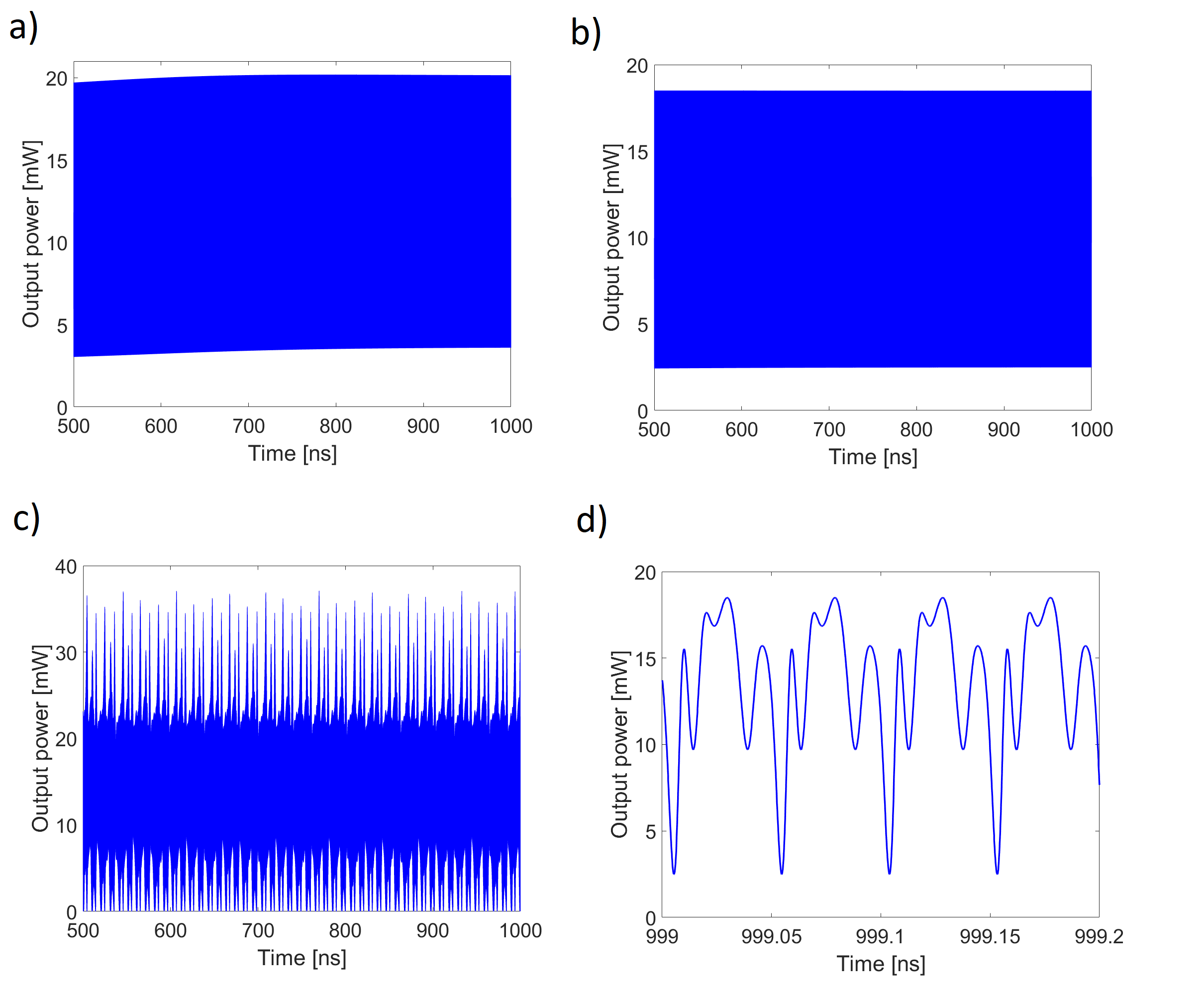}
\caption{$I_0=$1.54$I_\mathrm{thr}$, $\delta_\mathrm{hom}=$0.32THz: temporal evolution of the power in the last 500ns of the simulation for a) $I_\mathrm{A}=$0.30$I_\mathrm{thr}$, b) $I_\mathrm{A}=$0.54$I_\mathrm{thr}$, and c) $I_\mathrm{A}$=$I_\mathrm{thr}$; d) zoom of figure b) on a time interval of 200ps.}
\label{entire_time}
\end{figure}
After extracting the modulation frequency for each value of the current, we perform a stream of dynamical simulations by integrating the ESMBEs with Eq. (\ref{Ient}), for a the set of values of $I_A$ between 0.08$I_\mathrm{thr}$ and $I_\mathrm{thr}$ with step $\Delta I_\mathrm{A}=$0.08$I_\mathrm{thr}$, and we calculate the indicators $\sigma_P$ and $\sigma_{\Delta\Phi}$ for each simulated regime in order to investigate the impact of the modulation of the current on the locking of the system at different values of modulation depth.\\
We present the case $I_0=$1.54$I_\mathrm{thr}$, which is shown in Fig.~\ref{RF_entire}. We can observe that between $I_A=$0.08$I_\mathrm{thr}$ and $I_A=$0.32$I_\mathrm{thr}$ the values of the indicators are higher respect to the free running case ($I_A=$0) and the system results unlocked. At this values of $I_\mathrm{A}$, in fact, we report the occurrence of slow amplitude modulations (see Fig.~\ref{entire_time}a) in the power trace, which prevent the regular and identical repetions of structure which we observed in Fig.~\ref{OFCdense}a. This is due to a residual difference between the considered value of the modulation frequency $\Omega_\mathrm{M}$ and the effective BN frequency of the laser. Between $I_A=$0.4$I_\mathrm{thr}$ and $I_\mathrm{A}=$0.56$I_\mathrm{thr}$ the values of the indicators decreases and the systems returns into a locked state, without evidence of amplitude modulations during the total time of the simulations, as shown in Fig.~\ref{entire_time}b. For $I_\mathrm{A}>$0.54$I_\mathrm{thr}$ we observe a jump of the indicators at values which correspond to unlocked state of the system and fast modulations occur in the power trace, as it can be noticed in Fig.~\ref{entire_time}c. We explain this behaviour by noticing that the considered value of $I_0$ is near the border of the OFC region with the unlocked region (see Fig.~\ref{currentscan}) in absence of modulation, and therefore, when we introduce high values of the modulation depth, the system accesses for a relevant time interval this unlocked region, with a clear influence on the degree of locking. In order to corroborate this explanation, we performed the same study at lower values of $I_0$, and we found analogous jump of the indicators occurring at higher values of $I_A$ as lower the value of $I_0$.\\
In Fig.~\ref{entire_time}d we show an example of locked case in presence of RF injection, which is a zoom over a 200ps time interval of the 500ns long power trace shown in Fig.~\ref{entire_time}b for $I_\mathrm{A}=0.54I_\mathrm{thr}$. Respect to the free-running case (Fig.~\ref{OFCdense}a), we observe that the introduction of the RF injection produces a slight change in the shape of the multi-peaked power structure repeating each roundtrip. Therefore still a locked regime is obtained (unless the occurrence of slow power modulations that are not detectable in the total time interval of our simulation) but we are not able to generate a sequence of pulses. This characteristic is common to all the simulated locked regimes in presence of RF injection, confirming that an implementation of the modulation of the current in the entire cavity do not allow to produce a repetition of single-peak shaped narrow pulses \cite{Wang15}.\\ We also report that by introducing a modulation at 2$f_\mathrm{BN}$ over the entire cavity, we are not able to halve the repetition time of the OFC, but we still obtain power structures propagating each RT, with slightly different shape respect to the free running operation and the modulation at $f_\mathrm{BN}$ cases.\\
Finally, we consider the spontaneously generated HFC presented in Fig.~\ref{HFC}, and we introduce the modulation of the bias current at 2$f_\mathrm{BN}$, i.e. the frequency corresponding to the first peak observable in the power spectrum in Fig.~\ref{HFC}d. As a result of this numerical experiment, we obtain a locked regime with several differences respect to the free running case: a peak at the first BN arises in the power spectrum, the periodicity of the structures in the power vs time trace becomes 1RT, instead of 2RT of the original HFC, and the fundamental spacing in the optical spectrum is now the FSR instead of 2FSR. This investigation shows that HFC regimes are particularly ''fragile'' respect to the implementation of the RF injection, in the sense that even a modulation at 2$f_\mathrm{BN}$ implies a transition into a dense state. 
\begin{figure}[t]
\centering
\includegraphics[width=1\textwidth]{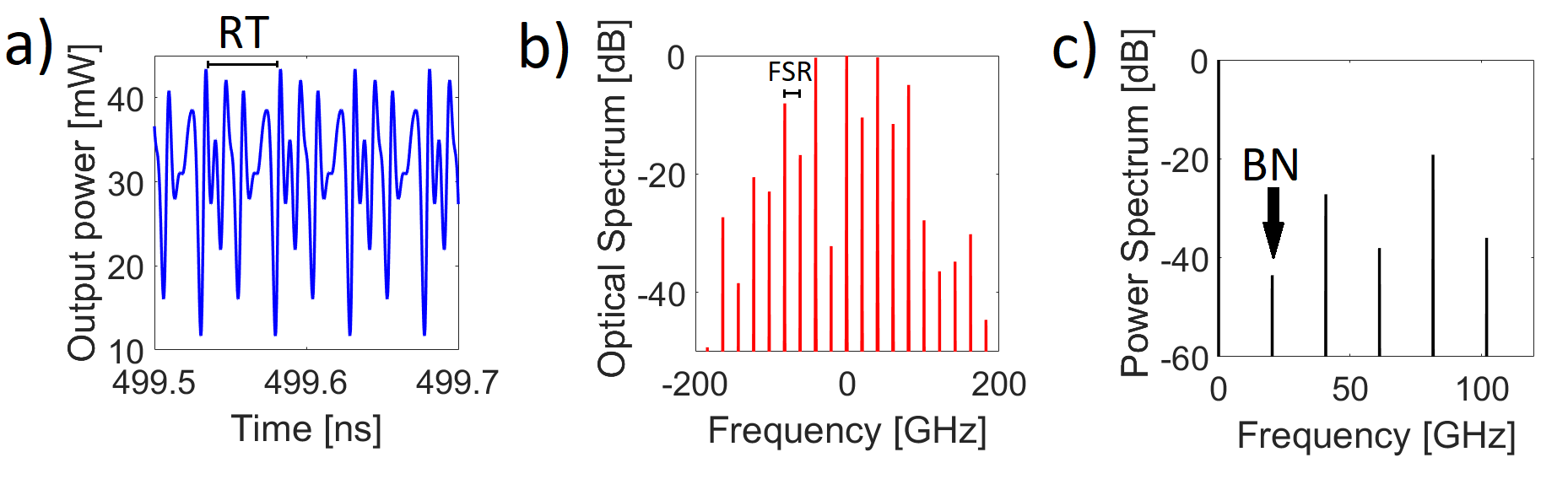}
\caption{Dense OFC regime obtained by introducing RF injection with $\Omega_M=2f_\mathrm{BN}$, for $I_0=2.15I_\mathrm{thr}$, $\delta_\mathrm{hom}=0.32\mathrm{THz}$, where in free running operation a 2nd order HFC was found. a) Temporal evolution of the power, b) optical spectrum, c) power spectrum. Other RF parameters: $p=1$,  $I_\mathrm{A}=0.5I_\mathrm{thr}$.}
\label{HFC_mod}
\end{figure}

\subsection{Generation of Pulses}
For generating a regular sequence of pulses it is necessary to introduce the modulation of the current into a section of the QCL cavity, by keeping the other portion in DC operation \cite{Wang15}. Therefore in this section we present the results obtained implementing the modulation of the current into the ESMBEs, by using  Eq.~\eqref{RFformula}, where a partial modulation of the laser cavity is described for $p>0$. 
\begin{figure}[!h]
\centering
\includegraphics[width=1\textwidth]{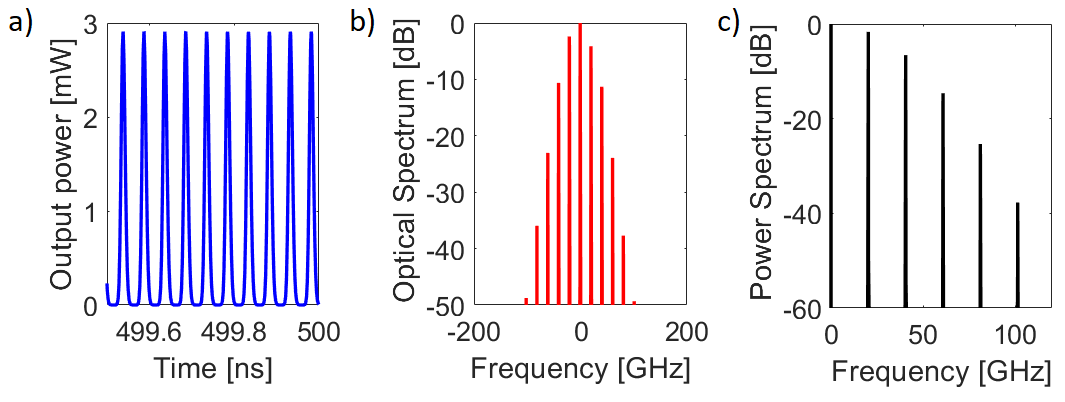}
\caption{Sequence of power pulses obtained for $\delta_\mathrm{hom}=0.32\mathrm{THz}$ and other parameters as in Table \ref{Table1}; $I_0=1.38I_\mathrm{thr}$, $I_A=0.31I_\mathrm{thr}$, $p=0.5$ (modulation of the current along half laser cavity), $I_B=0.62I_\mathrm{thr}$, $\Omega_M$ is the BN frequency extracted from the free running simulation at bias current $I_0$. a) Temporal evolution of the power; b) optical spectrum; c) power spectrum.}
\label{pulses}
\end{figure}
\begin{figure}[!h]
\centering
\includegraphics[width=0.9\textwidth]{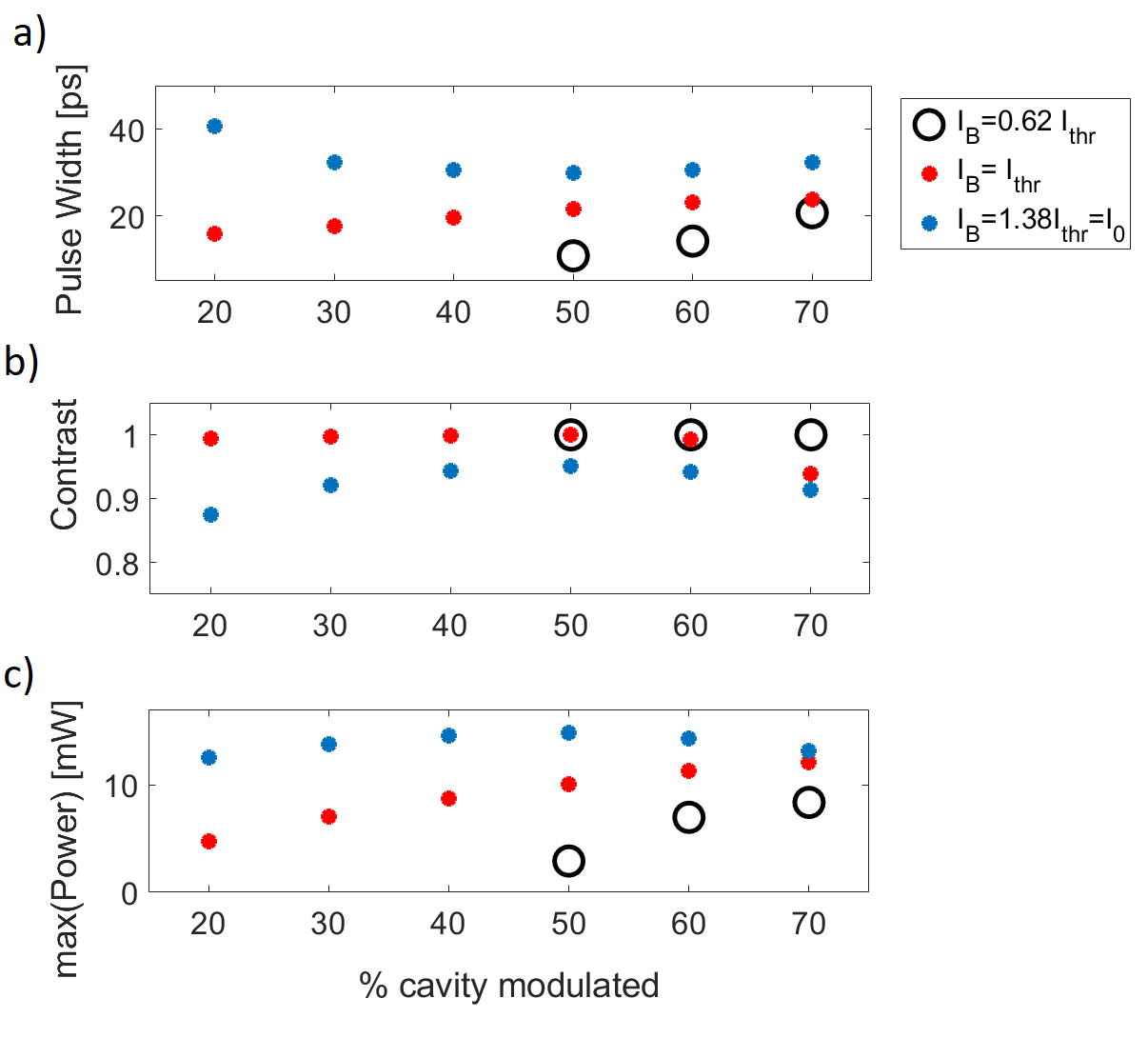}
\caption{Pulse properties for different values of the modulated portion of the laser cavity: a) pulse width, b) contrast, c) maximum power. The study has been repeated for three different values of the DC current in the unmodulated part: $I_\mathrm{B}=0.62I_\mathrm{thr}$ (black circle), $I_\mathrm{B}=I_\mathrm{thr}$ (red markers), and $I_\mathrm{B}=1.38I_\mathrm{thr}$ (blue markers).}
\label{pIb}
\end{figure}
An example of simulated regular sequence of pulses is shown in Fig.~\ref{pulses}a, obtained by introducing the RF modulation in half cavity ($p=0.5$), while in the other half the bias current is kept at constant value $I_\mathrm{B}=0.62I_\mathrm{thr}$, with a repetition rate corresponding to the FSR of the QCL (see Figs.~\ref{pulses}b--c) and a pulse FWHM of 10.8ps. For this case the central value of the current in the modulated portion is $I_0=1.38I_\mathrm{thr}$, and the modulation frequency $\Omega_M$ corresponds to the BN frequency extracted from the free running case at $I$=$I_0$.\\
\begin{figure}[t]
\centering
\includegraphics[width=1\textwidth]{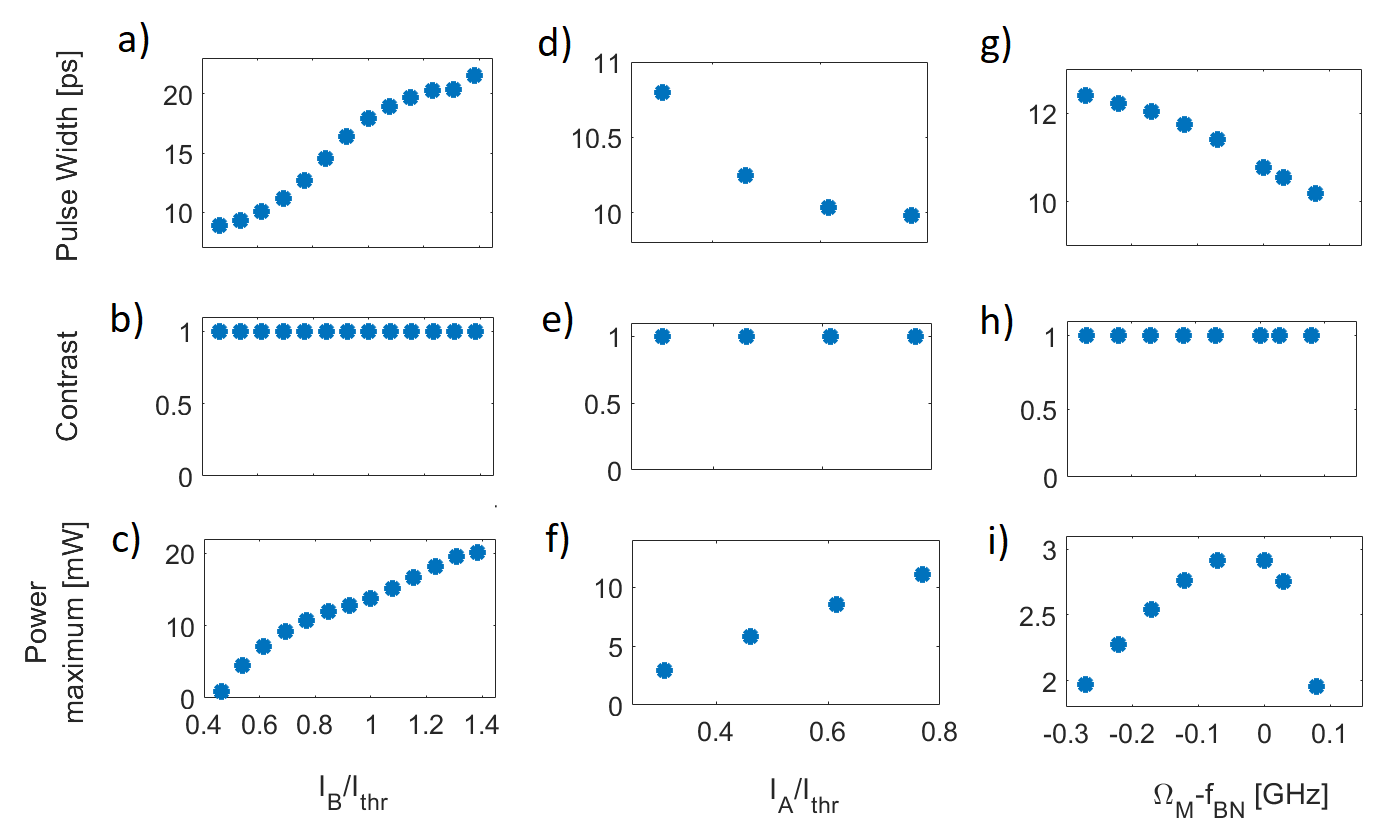}
\caption{Left column, role of $I_\mathrm{B}$: a) pulse width, b) contrast, c) maximum power as a function of $I_\mathrm{B}$ for fixed values $p=0.5$, $\Omega_\mathrm{M}=f_\mathrm{BN}$, $I_0=1.38I_\mathrm{thr}$,  and $I_\mathrm{A}=0.31I_\mathrm{thr}$.\\ Central column, role of $I_\mathrm{A}$: d) pulse width, e) contrast, f) maximum power as a function of $I_\mathrm{A}$ for fixed quantities $p=0.5$, $\Omega_\mathrm{M}=f_\mathrm{BN}$, $I_0$=$1.38I_\mathrm{thr}$, and $I_\mathrm{B}=0.62I_\mathrm{thr}$.\\ Right column, role of $\Omega_\mathrm{M}$: g) pulse width, h) contrast, i) maximum power as a function of the detuning $\Omega_\mathrm{M}-f_\mathrm{BN}$ for fixed quantities $p=0.5$, $I_0=1.38I_\mathrm{thr}$, $I_\mathrm{B}=0.62I_\mathrm{thr}$, and $I_\mathrm{A}=0.31I_\mathrm{thr}$.}
\label{complet}
\end{figure}
In order to better understand how to optimize the generation of pulses in this configuration, we perform a numerical study aimed to highlight the role of the modulation parameters $p$, $I_\mathrm{B}$, $I_\mathrm{A}$ and $\Omega_\mathrm{M}$ on the formation of a regular active-mode locked regime of short propagating pulses.\\
Firstly, we want to define the impact of the parameter $p$, i.e.~the length portion of the laser cavity subjected to RF modulation, on the characteristics of the generated pulses. In particular we want to explore how the pulse width, the maximum power, and the contrast vary with the parameter $p$ and how the scenario changes if we also vary $I_\mathrm{B}$. In this work the contrast is defined as $S=\frac{\mathrm{max}(\mathrm{Power})-\mathrm{min}(\mathrm{Power})}{\mathrm{max}(\mathrm{Power})}$.\\
For this purpose we set as a central value of the current in the modulated section  $I_0=1.38I_\mathrm{thr}$, we fix the modulation depth $I_\mathrm{A}=0.31I_\mathrm{thr}$, and we solve the ESMBEs for different values of $p$. We repeat this study for three different values of $I_\mathrm{B}$, choosing respectively a value below threshold ($I_\mathrm{B}=0.62I_\mathrm{thr}$), equals to threshold ($I_\mathrm{B}=I_\mathrm{thr}$), and above threshold ($I_\mathrm{B}=1.38I_\mathrm{thr}$).\\
Let us first examine the scenario when $I_\mathrm{B}$ is below threshold, which is described by the black circles in the three panels of Fig.~\ref{pIb}. For $p<0.5$ the emitted power vanishes, and therefore no data are reported in the figure. This is due to the fact that the current is below threshold for a portion bigger than half cavity, and the QCL is not able to provide a substantial emission of power. For $p \geq0.5$,  the contrast is 1 for all the obtained results, while the peak power of the pulses grows with $p$. In fact, an increase of $p$ implies a reduction in the length of section below threshold, and the system can provide pulses with higher maximum power. Furthermore, we remark that the lowest pulse FWHM is obtained for $p=$0.5, when half cavity is modulated: a value of 10.8ps is achieved.\\
If we set $I_\mathrm{B}$ to the threshold current (red markers in Fig.~\ref{pIb}), we report a general increase of the power for all the value of $p$ (Fig.~\ref{pIb}c), and at the same time also a general increase of the pulse FWHM (Fig.~\ref{pIb}a). We understand, therefore, that the value of $I_\mathrm{B}$ is crucial in order to obtain shorter pulses: the lower $I_\mathrm{B}$ is, the shorter pulses we obtain (see also Fig.~\ref{complet}a), with the negative effect that also the peak power decreases, as shown in Fig.~\ref{complet}c. This case, also, shows us that the pulse duration can be reduced by reducing the length of the modulated portion of the cavity (Fig.~\ref{pIb}a, red markers).\\
Furthermore, we can notice that the contrast (Fig.~\ref{pIb}b) decreases when we increase the length of the modulated section: here, the situation approaches the case of modulation in the entire cavity, where we have a reshaping of the structures occurring in the free running case, without generation of pulses. Conversely, for low values of $p$ we report contrast 1. \\If we furtherly increase $I_\mathrm{B}$ to the central current $I_0$ (blue markers in the three panels of Fig.~\ref{pIb}), for each value of $p$ the performances in terms of contrast and width get worst: we are not able to generate proper pulses, and only a variation in the shape of the free running structures is observed.\\
From the study presented in Fig.~\ref{pIb} it emerges that the configuration with contrast 1 and lowest pulse width is achieved for $p=0.5$ and $I_\mathrm{B}=0.62I_\mathrm{thr}$. If we keep fixed these values for these two quantities, we can observe that the scenario can be improved by increasing the modulation depth $I_A$.
In fact, as $I_\mathrm{A}$ increases, the peak power increases (see Fig.~\ref{complet}f) and we report a slight reduction of the pulse width (Fig.~\ref{complet}d): therefore it is possible to have shorter pulses with higher maximum power, and contrast 1 (Fig.~\ref{complet}e), by modulating with higher amplitude.\\
Finally, we want to investigate the role of the modulation frequency $\Omega_\mathrm{M}$. A modulation of the current with a slight detuning respect to the BN frequency $f_\mathrm{BN}$ extracted from the free running simulation, implies a modification of the characteristics of the pulses. In particular it is possible to reduce the pulse FWHM from 10.8ps (0 detuning) to a value of 10ps by an increase of 0.1 GHz to the modulation frequency $\Omega_\mathrm{M}$ with respect to $f_\mathrm{BN}$, as shown in Fig.~\ref{complet}g, with a concomitant reduction of the pulse peak power (Fig.~\ref{complet}i), which occurs in all the considered cases of modulation out of the resonance. We report contrast 1 for all the considered values of detuning, as presented in  Fig.~\ref{complet}b.
\subsection{Harmonic Mode-Locking}
A possible method to achieve the stable propagation of pulses of shorter duration, pushing the limit from the 10ps scale to the ps scale in the THz region, consists in the modulation of the current at a multiple of the BN frequency. Recently, several groups have experimentally implemented this technique, named harmonic mode-locking (HML), for the generation of pulses with THz QCLs, and new results have been presented \cite{Dhillon1}, \cite{Dhillon2}, \cite{Scalari1}. By adopting our model, we are able to reproduce numerically regular sequences of pulses by modulating the bias current at multiples $n$ of the BN frequency $f_\mathrm{BN}$.\\
\begin{figure}[p]
\centering
\includegraphics[width=1\textwidth]{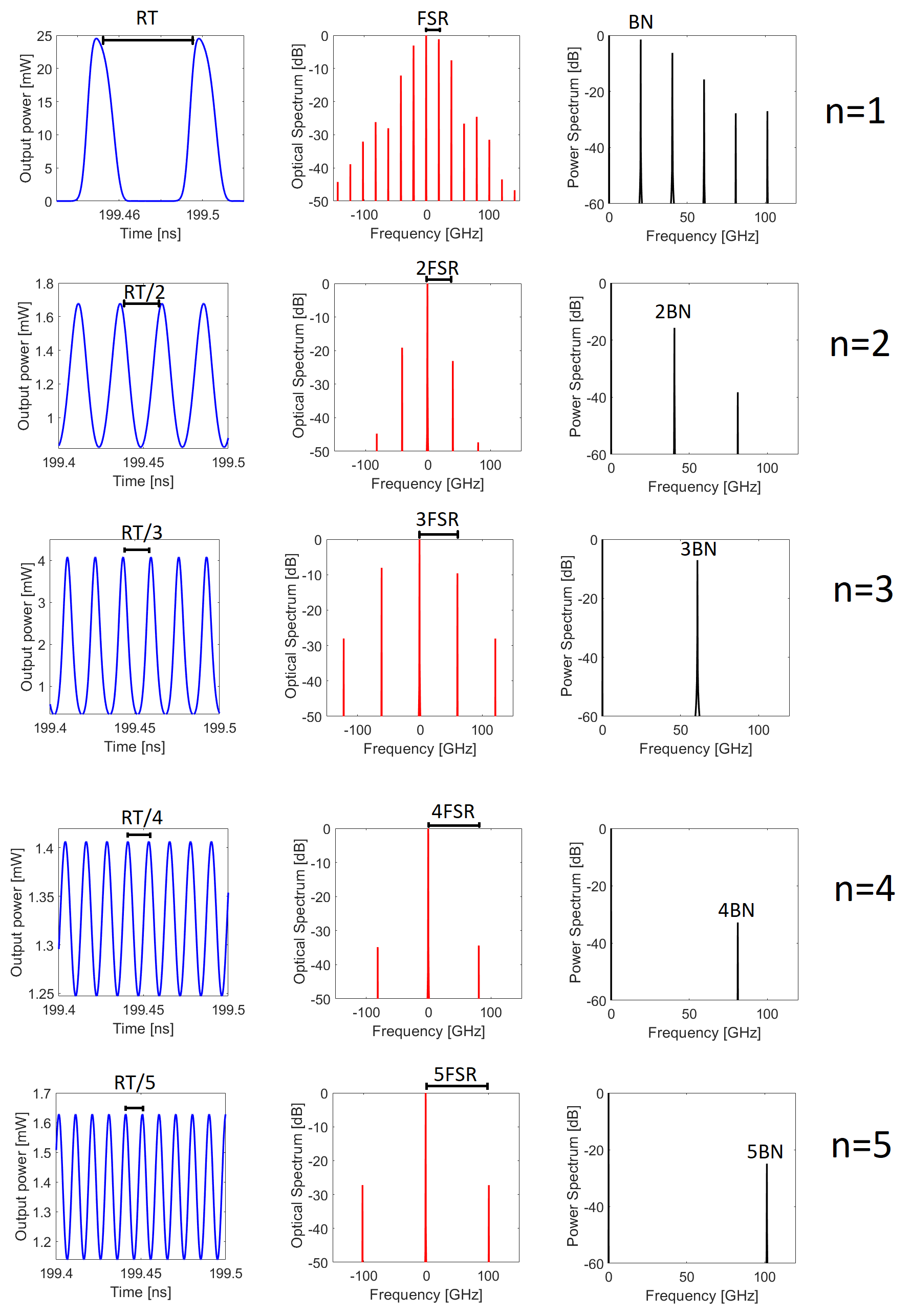}
\caption{Harmonic mode-locking for $p=0.5$, $I_\mathrm{B}=0.62I_\mathrm{thr}$, $I_\mathrm{A}=I_0=1.54I_\mathrm{thr}$: output power, optical spectrum and power spectrum for different order of harmonics $n$=1,...,5, where the modulation frequency is $\Omega_\mathrm{M}=$n$f_\mathrm{BN}$, and $f_\mathrm{BN}$ is the beat note frequency extracted from the free running simulation for a value of bias current ($I=I_0$).}
\label{HMLfig}
\end{figure}
In Fig.~\ref{HMLfig} we show our numerical results, related to order of harmonics $n$=1,...,5, obtained by modulating half cavity ($p=0.5$), and fixing $I_\mathrm{B}=0.62I_\mathrm{thr}$, $I_\mathrm{A}=I_0=1.54I_\mathrm{thr}$. For $n=1$ we have pulses propagating each roundtrip time RT and, by increasing $n$, the time interval between two consecutive pulses becomes RT/n. At the same time, the spacing between two adjacent modes in the optical spectrum increases from the FSR of the laser cavity to nFSR, and the first observable peak in the power spectrum is at n$f_\mathrm{BN}$. Furthermore, the suppression of modes in the optical spectrum when $n$ increases, implies a decrease in the peak power of the pulses and also the contrast is characterized by a reduction respect to the $n=1$ case, because the distance between the power of the side lines in the optical spectrum and the power of the central mode, tends to increase, and the emission of the QCL tends to resemble a CW behaviour with a slight superimposed modulation. This is clearly observable at $n=4$ and $n=5$. This effect would be mitigated by considering a larger gain spectrum, or a shorter FSR, so that a higher number of modes would access the portion of the optical spectrum nearby the peak, and the suppression of intermediate optical lines would not affect relevantly the contrast as in the shown cases for $n=4$ and $n=5$. 
\begin{figure}[t]
\centering
\includegraphics[width=0.8\textwidth]{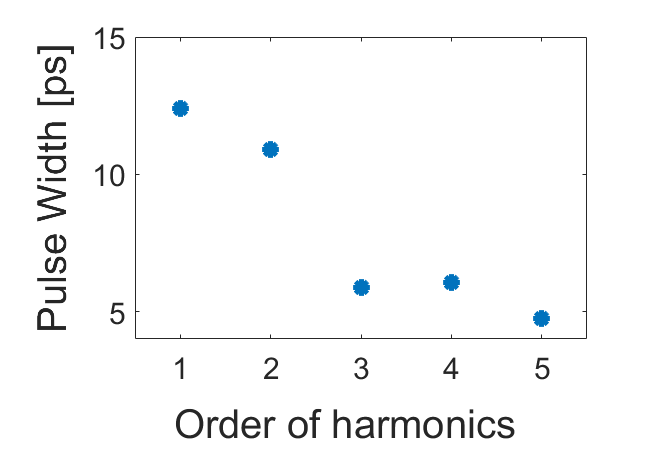}
\caption{Harmonic mode-locking for $p=$0.5, $I_\mathrm{B}=$0.62$I_\mathrm{thr}$, $I_\mathrm{A}=I_0=1.54I_\mathrm{thr}$: evolution of pulse FWHM with the order of harmonics $n$.}
\label{pulsewidthHML}
\end{figure}
If we examine the pulse duration for the different values of $n$, we can notice that the implementation of HML allows to switch from an initial value of 12ps obtained for $n=1$, to about 5 ps for $n=3$, as shown in Fig.~\ref{pulsewidthHML}, with the possibility to reach a value of 4.7ps for $n=5$, where however the performances in terms of contrast and maximum power reduce because of the already mentioned reasons. The slight increase of the pulse duration passing from $n=3$ to $n=4$, where the reverse would be expected because of a reduction of the repetition time, is a clear fingerprint of the passage from a regime with narrow pulses with contrast around 1, to a sinusoidal superimposed modulation of a CW emission. 

\section{Conclusions}\label{conclusion}
We performed an extended numerical study of the emission of dense and harmonic coherent regimes in THz-QCLs, exploring both the spontaneous generation and the active mode-locking. With our model, we are able to reproduce all the main experimental features characterizing the self-starting OFCs and HFCs in the THz range, remarking on the distinguishing points with respect to what is found in the mid-IR. In particular, we reproduced correctly the experimentally observed AM and FM behaviour of these spontaneous coherent regimes, such as the propagation of power multi-peaked structures and the absence of proper linear chirp of the instantaneous frequency; we observed pulsations of the density of carriers; we reproduced the spontaneous generation of harmonic regimes.\\
In the second part of the work, we investigated how the spontaneously generated combs are affected by the RF injection implemented in the entire cavity, identifying three different regimes of coherence for different ranges of the modulation depth. Then, we analyzed the case of partially modulated cavity, and we pushed our study to a systematic analysis of the impact of the modulation parameters on the duration, contrast, and peak power of the generated pulses. Finally, we performed the first numerical study about the harmonic mode locking, reproducing the reduction of repetition time and duration of the pulses provided by this technique.
\section*{Acknowledgements}
The authors acknowledge the funding from the Australian Research Council Discovery Project (Grant No. DP200101948). X.Q. acknowledges support from the Advance Queensland Industry Research Fellowships program.

\bibliography{sample}






\end{document}